\documentstyle[11pt,paspconf,psfig]{article}

\def\mincir{\raise -2.truept\hbox{\rlap{\hbox{$\sim$}}\raise5.truept
\hbox{$<$}\ }}
\def\magcir{\raise -2.truept\hbox{\rlap{\hbox{$\sim$}}\raise5.truept
\hbox{$>$}\ }}

\begin{document}

\title{Dark Matter Halos around Galaxies}
\author{P. Salucci}
\affil{SISSA, via Beirut 2-4, I-34013 Trieste, Italy}
\author{M. Persic}
\affil{Osservatorio Astronomico di Trieste, via Beirut 2-4, 
I-34013 Trieste, Italy}

\begin{abstract} 

We present  evidence that all galaxies, of any Hubble type and 
luminosity, bear the kinematical signature of a mass component
distributed differently from the luminous matter.  We review and/or derive
the DM halo properties of galaxies of different morphologies:  spirals,
LSBs, ellipticals, dwarf irregulars and dwarf spheroidals. 

We show that the halo density profile
$$
M_h(x) ~=~ M_h(1) ~(1+a^2) ~{x^3\over x^2+a^2}
$$ 
(with $x\equiv R/R_{opt}$), across both the Hubble and luminosity
sequences, matches all the available data that include, for ellipticals:
properties of the X-ray emitting gas and the kinematics of planetary
nebulae, stars, and HI disks; for spirals, LSBs and dIrr's: stellar and HI
rotation curves; and, finally, for dSph's the motions of individual stars. 

The dark + luminous mass structure is obtained: {\it (a)} in spirals, LSBs,
and dIrr's by modelling the extraordinary properties of the Universal
Rotation Curve (URC), to which all these types conform (i.e. 
the URC  luminosity dependence and the  smallness  
 of its rms scatter and cosmic variance);  {\it (b)} in ellipticals and dSph's, by
modelling the coadded mass profiles (or the $M/L$ ratios) in terms of a
luminous spheroid and the above-specified dark halo. 

A main feature of galactic structure is that the dark and  visible
matter are well mixed already in the luminous region. The transition
between the inner, star-dominated regions and the outer, halo-dominated
region, moves progressively inwards with decreasing luminosity, to the
extent that very-low-$L$ stellar systems (disks or spheroids) are not
self-gravitating, while in high-$L$ systems the dark matter becomes a   main 
mass component only beyond the optical edge. 

A halo core radius, comparable to the optical radius, is detected at all
luminosities and for all morphologies.  The luminous mass fraction varies
with luminosity in a fashion common to all galaxy types: it is comparable
with the cosmological baryon fraction at $L > L_*$ but it decreases by
more than a factor $10^2$ at $L << L_*$. 

For each Hubble type, the central halo density increases with decreasing
luminosity:  sequences of denser stellar systems (dwarfs, ellipticals,
HSBs, LSBs in decreasing order) correspond in turn to sequences of denser
halos. 

Then, the dark halo structure of galaxies fits into a well ordered pattern
underlying a unified picture for the mass distribution of galaxies across
the Hubble sequence.

\end{abstract}


\section{Introduction}

\begin{figure}
\par
\centerline{\psfig{figure=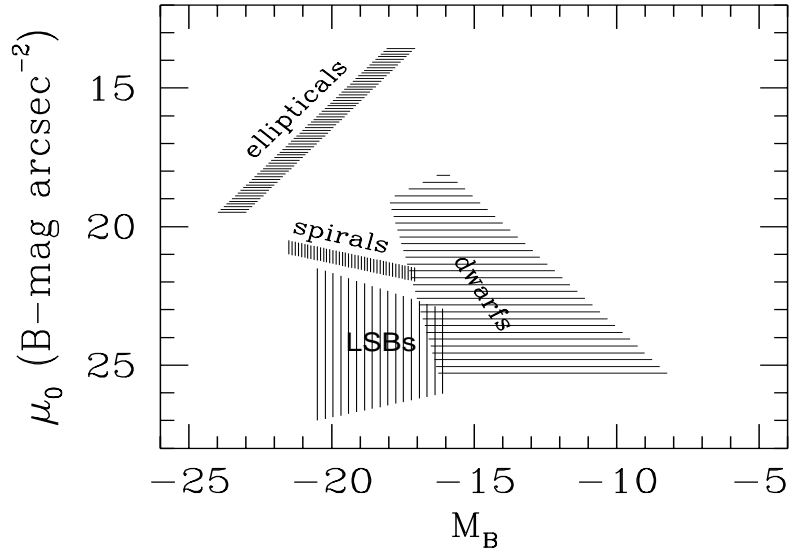,height=8.4cm,width=15.6cm}}
\par
\caption {The loci populated by the various families of 
galaxies in the central brightness vs. luminosity plane. }
\label{fig:MU_B}
\end{figure}

Over scales $\sim$ 1 kpc up to the Hubble radius, the
dynamics of cosmological systems is influenced, and often dominated, by
non-radiating matter which reveals itself only through a gravitational
interaction with the luminous matter. As the observational evidence has
been accumulating, it has become apparent that understanding the nature,
history and structural properties of this dark component, is the focal
point of Cosmology at the end of the millennium. In particular, the halos
of dark matter, detected around galaxies, have driven the dissipative
infall of baryons that, modulo a variety of initial conditions, has built
the bulge/disk/spheroid systems we observe today. It is remarkable that
the $\sim 10^{11}$ galaxies present within the Hubble radius can be
classified in a very small number of types: ellipticals, spirals,
low-surface-brightness (LSB) galaxies, dwarf spirals and dwarf irregulars.
The main characterizing property of these families is their position in
the $\mu_0$, M$_B$ (central surface brightness, magnitude) plane (see
Fig.~\ref{fig:MU_B}). In this plane spiral galaxies lie at the center and
show a very small range ($\sim 0.5$ mag) in $\mu_0$; ellipticals are very
bright systems, and span only a factor 10 in luminosity, that however well
correlates with central brightness; LSB galaxies are the counterpart of
spirals at low surface brightness; and, finally, dwarfs are
very-low-luminosity spheroids or disks which barely join the faintest
normal systems and span the largest interval in $\mu_0$.  Cosmologically,
all galaxy types are equally important for at least two reasons: {\it (a)}
those types having a lower average luminosity are however much more
numerous and hence can store a significant amount of baryons, and {\it
(b)} the properties that characterize and differentiate the various Hubble
types, i.e. the angular momentum content and the stellar populations, are
intimately related to the process of galaxy formation. 
 
A systematic presence of dark matter was first found in spirals,
specifically from the non-keplerian shape of their rotation curves (Rubin
et al. 1980; Bosma 1981), and in dSph's from their very high tidal $M/L$
ratios (Faber \& Lin 1983). Later, dark matter has been sistematically
found also around dwarf spirals and LSB galaxies (Romanishin et al. 
1982). For ellipticals, the situation is less clear: certainly at least
some (if not all) show evidence of a massive dark halo in which 
the luminous spheroid is embedded (e.g. Fabricant \& Gorenstein 1983).
Therefore, the claim for the ubiquitous presence of dark halos around
galaxies may be observationally supported. 
 
Theoretically, this claim is the natural outcome of the bottom-up
cosmological scenarios in which galaxies form inside dark matter halos
(probably with a universal density profile; e.g., Frenk et al. 1988 
and Navarro et al. 1996).  We point out that this prediction has so far
been tested only in spirals where reliable DM profiles have been obtained
(Persic, Salucci \& Stel 1996; hereafter PSS96). Along the Hubble
sequence, the systematic presence of dark matter in galaxies and its 
relation with
the luminous matter has so far been poorly known.  However, in the past
year or so a number of observational breakthroughs (some of which
presented at this conference) have allowed us to obtain the gravitational
potential in numerous galaxies of different Hubble types (including also
LSB galaxies). 

The time is now ripe for attempting to derive the general mass profile of
dark matter halos, as a function of galaxy luminosity and morphology.  In
this paper (which is also a review describing much recent work), we will
try to answer, for the first time, a simple (cosmological) question: 
Given a galaxy of Hubble type T and luminosity $L$, which halo is it
embedded in? 

The aim of this article is then to derive/review, by means of proper mass
modelling, the mass distribution in galaxies of different luminosities and
Hubble Types and to fit all of the various pieces into one unified scheme of
galaxy structure. Notice that the theoretical implications of the results
presented here will be discussed elsewhere. In detail, the plan of the
paper is as follows: in section 2 we review the properties of DM halos in
spiral galaxies; in section 3 we work out the DM mass distribution in LSBs
and perform a comparative analysis with that in spirals; in section 4 we
derive/review the halo properties of elliptical galaxies; in sections 5
and 6 we derive the properties of dwarf galaxies, irregulars and
spheroidals. Finally, in section 7 we propose a unified scheme for the DM
halos around galaxies and their interaction with the luminous matter. (A
value of H$_0=75$ km s$^{-1}$ Mpc$^{-1}$ is assumed throughout.)

\section{Spiral Galaxies}

The luminous ($\sim$stellar) matter in spiral galaxies is distributed in
two components: a concentrated, spheroidal bulge, with projected density
distribution approximately described by 
\begin{equation} 
I(R) ~=~ I_0 e^{-7.67\,(R/r_e)^{1/4}} 
\label{eq:DEVAUCOULEURS} 
\end{equation} 
(with $r_e$ being  the half-light radius; but see Broeils \& Courteau 1997), and
an extended thin disk with surface luminosity distribution very well
described (see Fig.\ref{fig:LIGHT_PROFILES}) by: 
\begin{equation} 
I(R) ~=~ I_0 ~ e^{-R/R_D}
\label{eq: FREEMAN} 
\end{equation} 
(with $R_D$ being the disk scale-length; Freeman 1970).  Let us take $R_{opt}$
as the radius encircling $83\%$ of the integrated light: for an
exponential disk $R_{opt}=3.2\,R_D$ is the limit of the stellar disk. The
relative importance of the two luminous components defines the Hubble
sequence of spirals, going from the bulge-dominated Sa galaxies to the
progressively more disk-dominated Sb/Sc/Sd galaxies.  We recall that the
spiral arms are non-axisymmetric density perturbations, traced by
newly-formed bright stars or HII regions, which are conspicuous in the
light distribution  and perturb the circular velocity field through small-amplitude
sinusoidal disturbances (i.e., wiggles in the rotation curves), but they
are immaterial to the axisymmetric gravitational potential. This
digression is to recall that fitting these features with a mass model is a
mistake! 

\begin{figure}
\par
\centerline{\psfig{figure=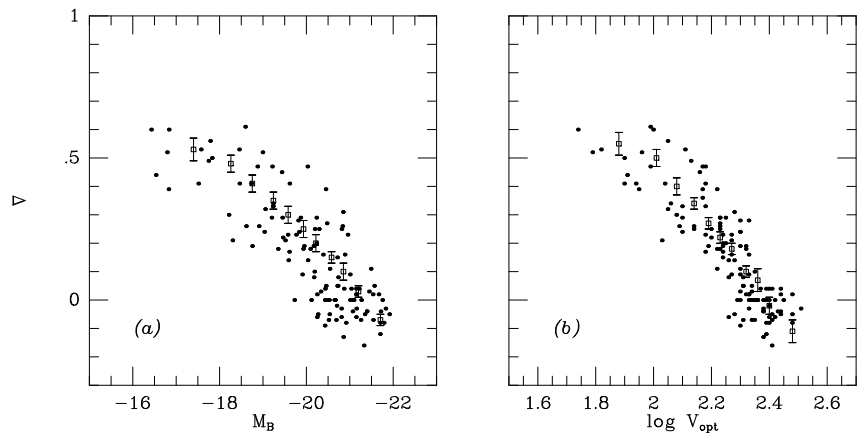,height=6.3cm,width=11.7cm,angle=-90}}
\par
\caption {The RC slope at $R_{opt}$ vs luminosity ({\it left}) and 
$V_{opt}$ ({\it right}) for the 1100 RCs of Persic \& Salucci 1995 and  
PSS96.} 
\label{fig:NABLA}
\end{figure}

\begin{figure}
\par
\centerline{\vbox{\psfig{figure=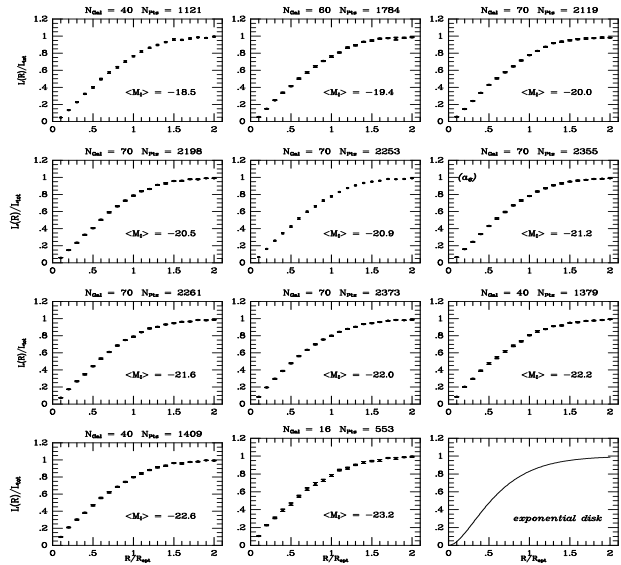,height=7.2cm,width=4.7cm} }}
\par
\caption { Averaged spirals  $I$-light profiles at different 
luminosities. Each $L$ bin    includes hundreths of galaxies} 
\label{fig:LIGHT_PROFILES}
\end{figure}

The rotation curves of spiral galaxies do not show any keplerian fall-off
at outer radii (e.g.: Rubin et al. 1980; Bosma 1981).  Moreover, their
shapes at $R>(1-2) R_D$ are inconsistent with the light distribution, so
unveiling the presence of a DM component. PSS96, analyzing approximately
1100 RCs, about 100 of which extended out to $\mincir 2\,R_{opt}$, found
that the luminosity specifies the entire axisymmetric rotation field of
spiral galaxies.  At any chosen normalized radius $x \equiv R/R_{opt}$,
both the RC amplitude and the local slope strongly correlate with the
galaxy luminosity (in particular, for $x=1$ see Fig.~\ref{fig:NABLA}; for
outer radii see PSS96, Salucci \& Frenk 1989, and Casertano \& van Gorkom 
1991).
Remarkably, the rms scatter around such relationships is much smaller than
the variation of slopes among galaxies (see PSS96). This has led to the
concept of the {\it Universal Rotation Curve} (URC) of spiral galaxies
(PSS96 and Persic \& Salucci 1991; see Fig.\ref{fig:URC}). The rotation
velocity of a galaxy of luminosity $L/L_*$ at a radius $x \equiv
R/R_{opt}$ is well described by: 
$$ 
V_{URC}(x) ~=~ V(R_{opt})~ \biggl[ \biggl(0.72+0.44\,{\rm log} {L \over
L_*}\biggr) ~{1.97~x^{1.22}\over{(x^2+0.78^2)^{1.43}}}~+ 
$$
\begin{equation} 
~~~~~~~~ +~ \biggl(0.28 - 0.44\, {\rm log} {L \over L_*}
\biggr) ~ \biggl[1+2.25\,\bigg({L\over L_*}\biggr)^{0.4}\biggr] ~ { x^2 \over
x^2+2.25\,({L\over L_*})^{0.4} } \biggr]^{1/2} ~~~~ {\rm km~s^{-1}}\,.
\label{eq:URC}
\end{equation} 
(with log$\,L_*/L_\odot=10.4$ in the $B$-band).  Remarkably, spirals show
a very small cosmic variance around the URC.  In $80\%$ of the cases the
difference between the individual RCs and the URC is smaller than the
observational errors, and in most of the remaining cases it is due to a
bulge not considered in eq.(\ref{eq:URC}) (Hendry et al. 1997; PSS96). This
result has been confirmed by a Principal Component Analysis study of URC
(Rhee 1996; Rhee \& van Albada 1997): they found that the two first
components alone account for $\sim 90 \%$ of the total variance of the RC
shapes. Thus, spirals sweep a narrow locus in the 
RC-profile/amplitude/luminosity space. 

\begin{figure}
\par
\centerline{\vbox{
\psfig{figure=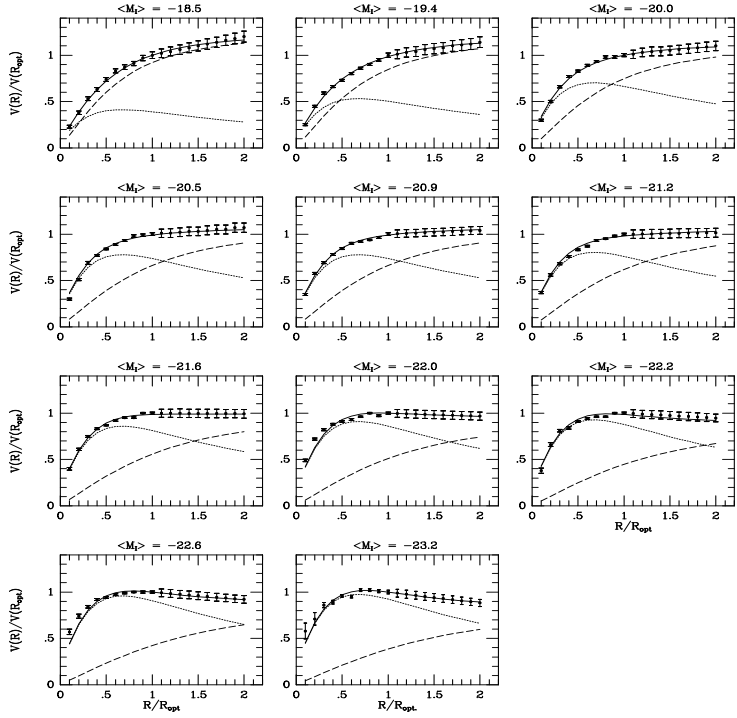,height=8.8cm,width=5.6cm} }}
\par
\caption { Coadded rotation curves (filled circles with error bars)
repruduced by URC (solid line) Also shown the separate dark/luminous  
contributions (dotted line: disk;  dashed line: halo.)}
\label{fig:VEL_PROFILES}
\end{figure}

\begin{figure}
\par
\centerline{\vbox{\psfig{figure=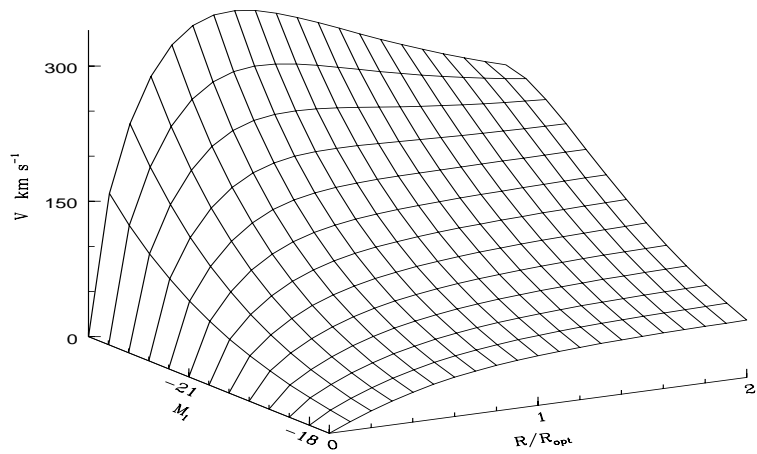,height=6cm,width=6cm} }}
\par
\caption{ The URC surface. }
\label{fig:URC}
\end{figure}

The luminosity dependence of the URC strongly contrasts with the
self-similarity of the luminosity distribution of stellar disks (Fig. 
~\ref{fig:LIGHT_PROFILES}): the luminosity profiles $L(x) \propto \int_0^x
x\,I(x)\, dx$ do not depend on luminosity.  This reflects the discrepancy
between the distribution of light and that of the gravitating mass.
Noticeably, this discrepancy increases with radius $x$ and with decreasing
galaxy luminosity $L$.  The URC can be fitted by a combination of two
components: {\it (a)} an exponential thin disk, approximated for $0.04 
R_{opt}< R \leq 2R_{opt}$ as
\begin{equation}
V_d^2(x)~ = ~V^2(R_{opt}) ~\beta~{1.97~ x^{1.22} \over (x^2+0.78^2)^{1.43}}\,,
\label{eq:DISK_VEL}
\end{equation}
and {\it (b)} a spherical halo represented by
\begin{equation}
V_h^2(x)~ = ~V^2(R_{opt}) ~(1-\beta) ~(1+a^2) {x^2\over{(x^2+a^2})}\,,
\label{eq:HALO_VEL}
\end{equation}
$$
M_h(x)=G^{-1} V^2(1) R_{opt}  ~(1-\beta) ~(1+a^2) {x^3\over{(x^2+a^2})}\,,
$$
with $x \equiv R/R_{opt}$ being the normalized galactocentric radius, $\beta
\equiv V_d^2(R_{opt})/V_{opt}^2$ the disk mass fraction at $R_{opt}$, and
$a$ the halo core radius (in units of $R_{opt}$).  The disk+halo fits to
the URC are extremely good (fitting errors are within 1\% on average) at all 
luminosities (see Fig.~\ref{fig:VEL_PROFILES}) when
\begin{equation}
\beta ~= ~0.72~ +~ 0.44\, {\rm log} \biggl({L \over L_*}\biggr)\,,
\label{eq:SPIRAL_BETA}
\end{equation}
\begin{equation}
a~ = ~1.5 \,  \biggl( {L \over L_*} \biggr)^{1/5}\,.
\label{eq:SPIRAL_CORE_RADIUS}
\end{equation}

Thus we detect, for the DM component, a central constant-density region of 
size $\sim R_{opt}$, slightly increasing with luminosity. The transition
between the inner, luminous-matter-dominated regime and the outer, 
DM-dominated regime
occurs well inside the optical radius: typically at $r << R_{opt}$ in
low-luminosity galaxies, and farther out, closer to $\sim R_{opt}$, at
high luminosities (see Fig.\ref{fig:VEL_PROFILES}). Thus, the ordinary
and dark matter are well mixed in the very stellar regions of spirals. 

The total halo mass can be evaluated by extrapolating the halo out to the
radius, $R_{200}$, encompassing a mean overdensity of 200.  We find:
\begin{equation}
R_{200} ~=~ 250~\biggl({L\over L_*}\biggr)^{0.2} {\rm kpc}
\label{eq:SPIRAL_R200}
\end{equation}
\begin{equation}
M_{200} ~\sim~ 2 \times 10^{12} \biggl({L \over L_*} \biggr)^{0.5}M_\odot\,,
\label{eq:SPIRAL_M200}
\end{equation} 
in good agreement with results derived from satellite and pair kinematics
(Charlton \& Salpeter 1991; Zaritsky et al. 1993).  Notice that
\begin{equation}
{M_{200} \over L_B} ~ \simeq ~ 75 \, \bigg({L_B \over L_*} \biggr)^{ -0.5}
\end{equation}
(in solar units).  This implies that the halo mass function is not
parallel to the observed galaxy luminosity function (Ashman, Salucci \& 
Persic 1993).
 
The luminosity dependence of the disk mass fraction can be interpreted in
terms of a mass-dependent efficiency in transforming the primordial gas
fraction into stars. In fact, the total (i.e., computed at $R_{200}$)
luminous mass fraction of a galaxy of luminosity $L$ is: 
\begin{equation}
{M_\star \over M_{200}} ~\simeq~ 0.05 ~ \biggl( {L\over L_*} \biggr)^{0.8}\,.
\end{equation}
This suggests that only the brightest objects reach a value comparable
with the primordial value $\Omega_{BBN} \mincir 0.10$, while in
low-$L$ galaxies only a small fraction of their original baryon
content has been turned into stars.  The luminosity dependence of the DM
fraction found by Persic \& Salucci (1988, 1990) has been confirmed by
direct mass modelling (e.g.: Broeils 1992; Broeils \& Courteau 1997; 
Sincotte \& Carignan 1997; see also Ashman 1992). 

On scales $(0.2-1)\,R_{200}$, the halos have mostly the same structure,
with a density profile very similar at all masses and an amplitude that
scales only very weakly with mass, like $V_{200} \propto M_{200}^{0.15}$. 
This explains why the kinematics of satellite galaxies orbiting around
spirals show velocities (relative to the primary) uncorrelated with the
primary's luminosity (Zaritsky 1997).  At very inner radii, however, the
self-similarity of the profile breaks down, the core radius becoming
smaller for decreasing $M_{200}$ according to: 
\begin{equation} 
{{\rm core ~ radius} \over R_{200}} ~=~ 0.075\, \biggl( {M_{200} \over 
10^{12} M_\odot} \biggr)^{0.6} \,.
\end{equation} 
The central density scales with mass as:  
\begin{equation} 
\rho_h(0) ~ =~ 6.3 \times 10^4
\rho_c \, \biggl( {M_{200} \over 10^{12} M_\odot} \biggr)^{-1.3} 
\end{equation}
($\rho_c$ is the critical density of the universe). 

The regularities of the luminous-to-dark mass structure can be represented
as a curve in the space defined by dark-to-luminous mass ratio,
(halo core radius)-to-(optical radius) ratio, central halo density (see
PSS96). This curve is a structural counterpart of the URC and represents
the only locus of the manifold where spiral galaxies can be found.  The
main properties of dark matter in spiral galaxies can then be summarized
as follows:  
\medskip

\noindent
$\Box$ substantial amounts of dark matter are detected in the optical
regions of all spirals, starting at smaller radii for lower luminosities; 
\medskip

\noindent
$\Box$ the dark and the luminous matter are well mixed; 
\medskip 

\noindent
$\Box$ the structure of the halos is universal: it involves a core
radius comparable in size with the optical radius and a central density
scaling inversely with luminosity; 
\medskip

\noindent
$\Box$ the ratio between luminous and dark matter is a function of
luminosity (or mass) among galaxies.  At a given (normalized) radius, this
ratio increses with increasing luminosity: the global visible-to-dark mass
ratio spans the range between $\sim \Omega_{BBN}$ at very high
luminosities and $\sim 10^{-4}$ at $L << L_*$.
\medskip 

The discovery of these features, describing the various stages of the
processes leading to present-day galaxies, supersedes the observationally
disproved paradigms of "flat rotation curves" and "cosmic conspiracy".

\section {Low-Surface-Brightness Galaxies}

The central surface luminosity of spirals is normally about constant, at
$\mu_0(B) = 21.65 \pm 0.30$. Recently, however, a large population of disk
systems with a significantly lower surface brigthness ($\mu_0(B) = 24 -
25$) has been detected (Schombert et al. 1992; Driver et al. 1994; 
Morshidi et al. 1997; see also Disney \& Phillips 1983). In these systems
the light distribution follows that of an exponential thin disk (McGough
\& Bothun 1994), with total magnitude $\sim 1.5$ mag fainter than that of
normal (HSB) spirals.  At the faint end of the LSB luminosity distribution
(M$_B \sim -16$), the galaxies have very extended HI disks whose kinematics
yields the mass structure (de Blok, McGaugh \& van der Hulst 1996).  In
detail, de Blok et al. have published the HI rotation curves of 19 low-$L$
LSBs, having M$_B \sim -17$ mag, $V_{max} \sim 70$ km 
s$^{-1}$, and (remarkably) $R_{opt} \sim (8-10)$ kpc, roughly
independent of luminosity. These galaxies are the counterpart of the
faintest HSB spirals. Notice that, although the maximum velocities are
similar, the optical sizes of LSBs are $\sim 3$ times larger than those of
HSBs. 

\begin{figure}
\par
\centerline{\vbox{
\psfig{figure=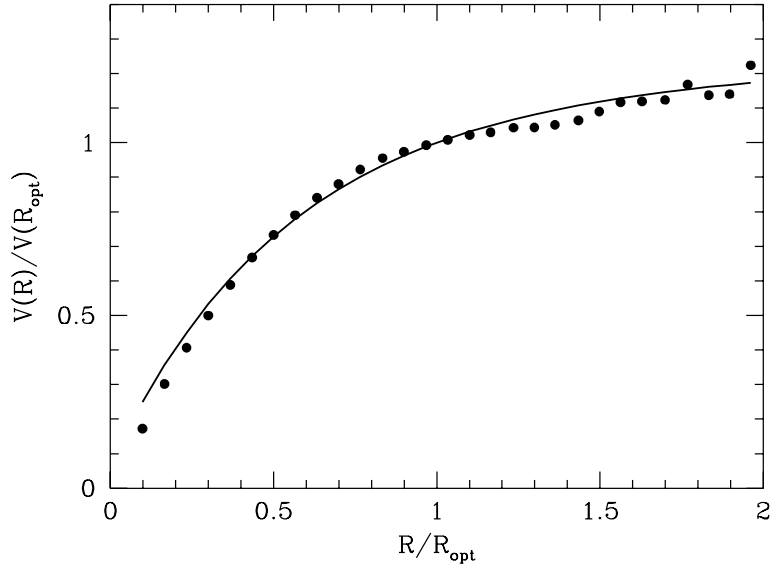,height=7.6cm,width=6cm} }}
\par
\caption { Coadded rotation curve of the sample of LSB galaxies with 
$V_{opt} \sim (70 \pm 30)$ km s$^{-1}$. The solid line represents the 
$V_{\rm URC}$ of spirals of similar $V_{opt}$. } 
\label{fig:LSB_CURVE}
\end{figure}

The objects in de Blok et al. (1996) are all within a small range of
magnitudes and optical velocities, $M_B=-17 \pm 0.5$, $V_{opt} \sim (70
\pm 30)$ km s$^{-1}$. From these data we construct, as we did for spirals,
the coadded rotation curve $V({R \over R_{opt}}, 70)$.  Notice that,
unlike for spirals, rotation curves of LSBs with higher $V_{opt}$ are not
available: so we can compare LSB and spiral RCs only at their lowest
velocities, shown for spirals in the top left panel in
Fig.\ref{fig:VEL_PROFILES} and for LSBs in Fig.\ref{fig:LSB_CURVE}
(points). In detail, in the latter figure we plot $V(R/R_{opt})$, the
coadded RC obtained from the de Blok et al. data, together with the spiral
$V_{\rm URC}$ for the same range of maximum velocities (50--100 km
s$^{-1}$) (see PSS96 for all of the details).  The agreement is striking: 
no difference can be detected between the LSB and HSB rotation curves,
which coincide within the observational uncertainties. Notice that the LSB
and HSB rotation curves are identical when the radial coordinate is
normalized to the disk length-scale (essential procedure for determining
the DM distribution).  The fact that, when expressed in physical radii,
LSB rotation curves rise to their maximum more gently than HSB RCs (see
Fig.3 of de Blok \& McGough 1997), depends exclusively on the larger LSB
disk scalelengths. 

For LSBs we adopt a mass model including, as for spirals, {\it (i)} an
exponential thin disk and {\it (ii)} a dark halo of mass profile given by
eq.(\ref{eq:HALO_VEL}).  The LSB coadded RC is extremely well fitted by
eqs.(\ref{eq:DISK_VEL}) and (\ref{eq:HALO_VEL}) with the `spiral' values
$\beta=0.1 \pm 0.04$ and $a \simeq 0.75$.  These parameters imply that LSB
galaxies are completely dominated by a dark halo with a large $\sim (5-6)$
kpc core radius. 

\begin{figure}
\par
\centerline{\hbox{
\psfig{figure=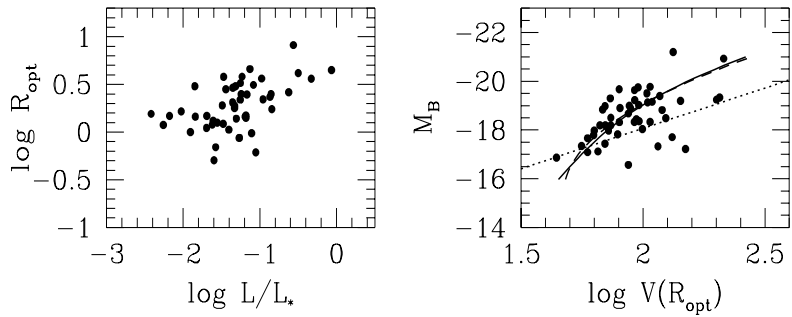,height=5cm,width=5cm} }}
\par
\caption { The radius--luminosity relation {\it (left)} and 
the Tully-Fisher relation {\it (right)} for the Matthews sample
of LSBs. Dashed curve: fit to the data. Solid curve: predictions 
of a mass model identical to that of spirals. 
  Dotted line: prediction of a mass model 
 model with constant $(M/L)$}
\label{fig:LSB_TF_RL} 
\end{figure}

Higher-luminosity (up to M$_B \sim -22$) LSBs do exist (e.g., Sprayberry
et al. 1993): their structure can be tentatively investigated by means of
their Tully-Fisher relationship. Observationally, these objects show a
good correlation between luminosity and the corrected linewidth $w_{0,i}$
(Matthews et al.  1997): 
\begin{equation}
{\rm log} w_{0,i} ~=~ 2.41 ~+~  0.69\, {\rm log} \biggl( {L_V \over L_V^*} 
\biggr) ~+~ 0.16 \, {\rm log}^2  \biggl( {L_V \over L_V^*} \biggr) \,,
\end{equation}
where $L_V^*$ corresponds to M$_V^*=-20.87$ (see Fig.~\ref{fig:LSB_TF_RL}).
The quantity $w_{0,i}$ is, as in spirals, a good measure of the
circular velocity at $\sim R_{opt}$: then  
\begin{equation}
w_{0,i}^2\, R_{opt}  ~\simeq~  G ~ [M_h(R_{opt})+ M_{bar}(R_{opt})]\,,
\end{equation}
where $M_{bar}=M_D+M_{HI}$. For LSB spirals we take the same
dark-to-visible mass ratios as for HSB spirals, ${ M_h(R_{opt}) \over
M_{bar}(R_{opt}) } \simeq 9 \times ( {L_B \over 0.04\, L_*} )^{-1}$ (see
conclusions of PSS96), and we assume $M_{bar}(R_{opt}) \propto L_B^{k}$.
Finally, by using the LSB luminosity-radius relation, $R_{opt} ~=~ 12 \,
(L/L_*)^{0.25}$ kpc (see Fig.~\ref{fig:LSB_TF_RL}), we are able to
reproduce the observed M$_V$--log$\,w_{0,i}$ relation: in
Fig.~\ref{fig:LSB_TF_RL} we show the excellent agreement between the
observed linewidths and those predicted by our mass model when $k =1.8$
and the LSBs stay on the `spiral' $\beta$--$L$ and $a$--$L$ relationships. 

\begin{figure}
\par
\centerline{\hbox{
\psfig{figure=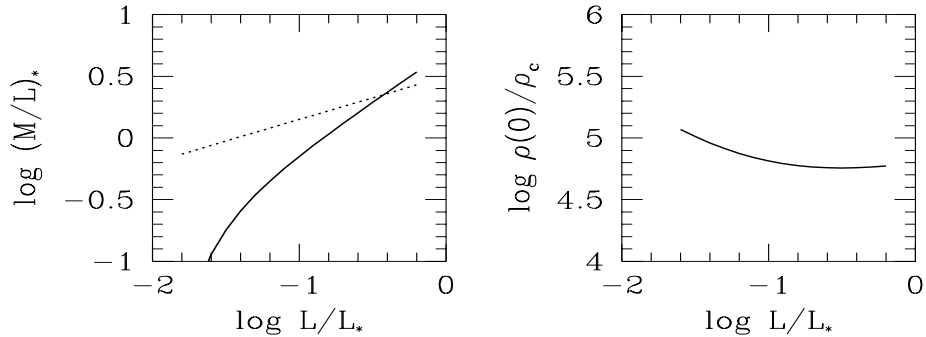,height=4.5cm,width=4.5cm} }}
\par
\caption { LSB $(M/L)_\star$ (left: solid curve; 
the dotted line refers to spirals) and central overdensity (right) as a 
function of luminosity.}  
\label{fig:LSB_MODEL} 
\end{figure}

At $\sim$$L_*$, the LSB stellar $M/L$ ratios are similar to those of HSBs,
but they decline steeply with decreasing luminosity (see
Fig.~\ref{fig:LSB_MODEL}, left panel). This result is in good agreement
with the $(M/L)_\star$ ratios obtained by applying the maximum-disk
hypothesis to solve for the galactic structure: in this case
log$\,(M/L)_\star$ increases from $\sim -0.3 $ to $0.5$ across the entire
LSB luminosity (velocity) range (de Blok \& McGaugh 1997). 

The central overdensities, $3/(4 \pi G \rho_c) (1-\beta) (V_{opt}/
R_{opt})^2 (1+a^2)/a^2$, are shown in Fig.~\ref{fig:LSB_MODEL} (right
panel) as a function of luminosity. They are smaller and less dependent on
galaxy luminosity than those in normal spirals, in agreement with the
findings by de Blok \& McGough (1997). 

With the caveat of the relative smallness of the present sample, we claim
that LSB galaxies are indistinguishable from HSBs in the $(\beta,\,a$)
space describing  the coupling between dark and visible matter. 

To end this section, we comment on the argument, raised sometimes,
according to which the discovery of a large number of LSBs would make the
findings on DM in spirals (e.g. PSS96) irrelevant, in that normal spirals
would represent an unfair, biased sample of galaxies. There is no doubt
that the existence of a population of LSB galaxies, contiguous to the
HSBs, affects the interpretation, and even the physical meanings, of
luminosity functions and number counts. However, it is exactly the
characteristic difference between the two families that answers the point. 
Even prior to any mass modelling, we can argue that in the Universe there
are about $10^{11}$ disk systems obeying the ``Freeman law'': $\mu_0= 21.5
\pm 0.5$ independent of galaxy luminosity. It is obviously necessary to
know the DM properties of this family, independently
of the existence of another family of disk systems not obeying such a
``law'', and maybe not even observable. Of course, investigating the DM
properties of LSBs is equally important and crucial.
Furthermore, the argument raised above appears even more
immaterial after the analysis of the LSB rotation curves. In fact, as far
as the structural parameters are concerned, HSBs and LSBs are
indistinguishable.  In other words, LSBs follow the spiral $V_{\rm
URC}(R/R_{opt})$, but strongly deviate with respect to the spirals'
luminosity--(disk length-scale) relation.

\section{Elliptical Galaxies}

Elliptical galaxies are pressure-supported, triaxial stellar systems whose
orbital structure may depend on their angular momentum content, degrees of
triaxiality and velocity dispersion anisotropy.  As is well known, the
derivation of the mass distribution from stellar motions is
not straightforward as it is from rotation curves in disk systems, because
the kinematics of the former is strongly affected by geometry, rotation, and
anisotropies. However, in addition to stellar kinematics, there are a
number of mass tracers (X-ray emitting halos, planetary nebulae, ionized
and neutral disks) which crucially help to probe the gravitational
potential out to external radii (see Danziger 1997). 

\begin{figure}
\par
\centerline{\vbox{\psfig{figure=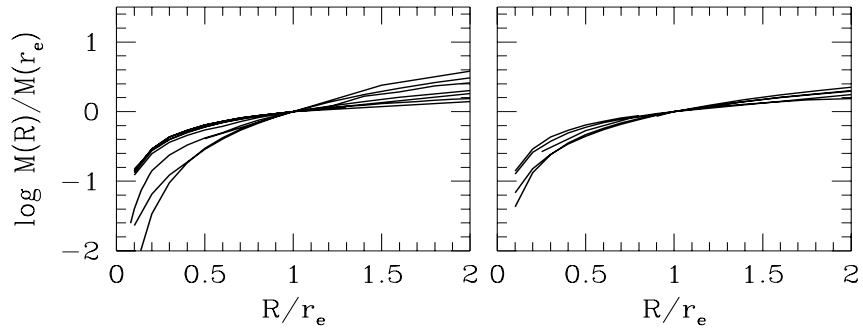,height=5cm,width=5cm}}}
\par
\caption { The mass as a function of radius for the
ellipticals of our subsamples, with luminosities $< {\rm log}\,L/L_*>=-0.4$ 
and $< {\rm log}\,L/L_*=0.4$ (left and right, respectively).}
\label{fig:ELL_MASS_PROF}
\end{figure}

\begin{figure}
\par
\centerline{\vbox{\psfig{figure=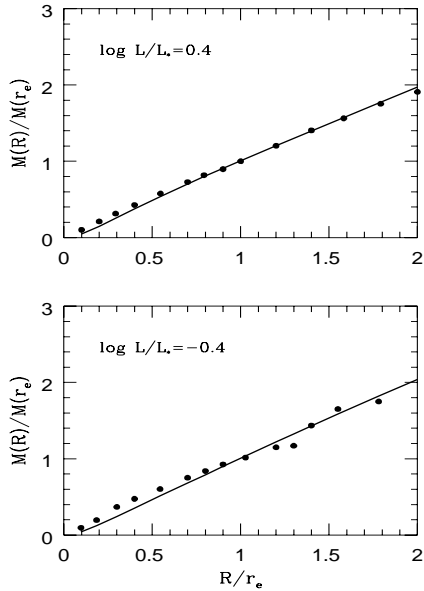,height=8cm,width=5cm}}}
\par
\caption {  Coadded mass profile of ellipticals,  filled cirles, with the
best fit mass model (solid line) }
\label{fig:ELL_MASS_COADDED}
\end{figure}

The luminosity profile $\rho_\star(r)$ of E's is obtained  from the observed
projected surface density, which follows the  de Vaucouleurs profile [see
eq.(\ref{eq:DEVAUCOULEURS})], by assuming an intrinsic shape 
and deprojecting. A very good approximation for $\rho_\star(r)$ 
is the Hernquist profile: 
\begin{equation}
\rho_\star (r) ~=~ {M_\star \over 2\pi}~ {1\over{ y (y+ c)^3} }
\label{eq:HERNQUIST}
\end{equation}
(where $y=1.81 \,r/r_e$ and $c=1$). As the Hernquist profile is no longer a 
good fit for $R >> r_e$, from actual surface-photometry profiles we have 
evaluated that $R_{opt} \simeq 2\,r_e$, where $r_e = 6\,(L/L_*)^{0.7}$ kpc 
(from Djorgovski \& Davis 1987). All ellipticals belong, within a 
very small cosmic scatter ($< 12\% $), to a relation of the type:
\begin{equation} 
r_e ~ \propto ~ \sigma_0^A ~ I_e^B \,, 
\label{eq:FP}
\end{equation} 
where $r_e$ is the half-light radius, $I_e$ is the mean 
surface brightness
within $r_e$, and $\sigma_0$ is the observed (projected) central velocity
dispersion.  In the logarithmic space, this corresponds to a plane, the
Fundamental Plane of ellipticals (Djorgovski \& Davis 1987; Dressler et
al. 1987), that constrains the properties of the DM distribution.  

Observations indicate $A=1.23$ ($1.66$) and $B=-0.82$ ($-0.75$) in the $V$-band and $K$-band,
respectively (e.g., Djorgovski \& Santiago 1993). Assuming a constant $M/L$ 
and structural homology, the 
virial theorem predicts $A=2$, $B=-1$. The simplest explanation of the
departure of $A$ from the virial expectation involves a systematic
variation of $M/L$ with $L$, which also accounts for the wavelength
dependence of $A$ (Djorgovski \& Santiago 1993). The departure of $B$ from 
the virial expectation is likely to be 
due to the breakdown of the homology of the luminosity structure 
(see Caon, Capaccioli \& D'Onofrio 1993; Graham \& Colless
1997). Assuming spherical symmetry and isotropic stellar motions 
(so $M_\star \simeq 3.4 G^{-1}\sigma_0^2r_e$ and $L=2\pi I_e r_e^2$), 
the FP implies that $M/L|_{r_e}$, inclusive of
dark matter, is ``low'': $4-9$ (Lanzoni 1994; Bender, Burstein \& 
Faber 1992), and roughly consistent with the values 
predicted by the stellar population models (Tinsley 1981). No large
amounts of DM are needed on these scales, as it also emerges from 
mass models of individual ellipticals, obtained by analyzing the line
profiles of the l.o.s velocity dispersion (van der Marel \& Franx 1993),
which show that the DM fraction inside $r_e$ is substantially less than 
$50\%$ (i.e.
$M/L_B \mincir 10$; see van der Marel 1991, 1994; Rix et al. 1997;  Saglia
et al. 1997a,b; Carollo \& Danziger 1994; Carollo et al.  1995), as in
spirals. 

We now investigate in some detail the effects of DM on the Fundamental
Plane. For this purpose, let us describe, for mathematical simplicity, the
dark halo by a Hernquist profile [eq.(\ref{eq:HERNQUIST})] with a lower
mass concentation than the luminous spheroid:  $c = 2$ (see Lanzoni 1994).
We recall that dark halos are likely to have an innermost
constant-density region not described by eq.(\ref{eq:HERNQUIST}): 
this, however, has no great relevance here, in that most of
the dark mass is located outside the core where eq.(\ref{eq:HERNQUIST})
is likely to hold. Without loss of generality, we consider isotropic
models (Lanzoni 1994): the radial dispersion velocity $\sigma_r$ is
related, through the Jeans equation, to the mass distribution by: 
\begin{equation} 
\sigma_r^2(r) ~=~ G\,{M_{dark}(r) ~+~
M_\star (r) \over r} ~ \biggl( {d{\rm log} \rho_\star \over d{\rm log}
r}\biggr)^{-1} \,.
\label{eq:SIGMA_R}
\end{equation} 
The projected velocity dispersion is then $\sigma_P(R) ~=~ {2 \over
\Sigma_\star(R)} \int_R^\infty { \rho_\star (r) \sigma^2_r(r) \over 
\sqrt{r^2-R^2}
} dr$. This equation shows that, at any radius (including $r=0$), the
measured projected velocity dispersion $\sigma_P(R)$ depends on the
distributions of both dark and luminous matter out to $r \sim (2-3)\, r_e$. 
Conversely, in spirals the circular velocity $V(R)$ depends essentially
only on the mass inside $R$, namely on just the luminous mass when $R
\rightarrow 0$. If $M_{200}$ is the total galaxy mass, we get
\begin{equation}
\sigma_P(0) ~=~ 3.4 ~{GM_\star \over r_e} ~ \biggl( {M_{200}\over 10 \, 
M_\star}\biggr)^{2/5}\,. 
\label{eq:SIGMA}
\end{equation}

The mass dependence of $\sigma_P(0)$, combined with the thinness of the FP
(i.e.: $\delta R_e/R_e \mincir 0.12$), constrains the scatter that would
arise, according to eq.(\ref{eq:SIGMA}), from random variations of the 
total amount of DM
mass in galaxies with the same luminous mass. From eq.(\ref{eq:FP}) we get
$\delta\sigma/\sigma \mincir {0.12 \over 1.4}$, while eq.(\ref{eq:SIGMA})
implies $d\sigma/\sigma = 5/2\, \delta M/M$. This means that, over 2
orders of magnitude in $M_\star$, any random variation of the dark 
mass must be less than $20\%$ (Lanzoni 1994;  Renzini \& Ciotti 1993;
Djorgovski \& Davis 1987).  From eq.(\ref{eq:FP}) and since $M_\star
\propto L^{1.2}$, we finally get ${M_\star\over M_{200}} ~ \propto ~
\sigma^{5/2}$, so that $M_{200} \propto L^{0.5-0.6}$, as in spirals (PSS96). 
Let us notice that the these well-established constraints on the amount of
dark matter in ellipticals, due to the existence of the Fundamental Plane,
are however at strong variance with the extremely high values of the
central $M/L$ ratios, $15-30$, found in some objects, as a result of the
dynamical models of Bertin et al. (1992) and Danziger (1997). 
 
\begin{figure}
\par
\centerline{\vbox{
\psfig{figure=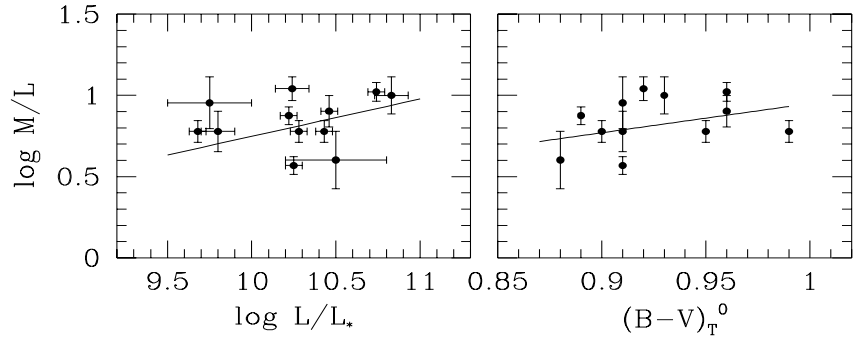,height=5cm,width=5cm} }}
\par
\caption { The $M/L$ ratio for a sample of ellipticals as a function of 
luminosity  (left) and color (right). The solid lines indicate 
the slopes of 0.2 and 1.8 (left and right, respectively).} 
\label{fig:ELL_ML}
\end{figure}

We can determine the parameters of the dark and visible matter
distribution by means of a variety of tracers of the ellipticals'
gravitational field (see the review by Danziger 1997). This provides the
(dark + luminous) mass distribution for 12 galaxies, with magnitudes
ranging between $-20 < {\rm M}_B < -23$. In order to investigate the
luminosity dependence of the mass distribution, we divide the sample into
2 subsamples, with average values $< {\rm log}\,L/L_*>= -0.4$ and $<{\rm
log}\,L/L_*> =0.4$ respectively.  In Fig.~\ref{fig:ELL_MASS_PROF} we plot,
as a function of $R/r_e$, the normalized mass profile of ellipticals,
$M(R)/M(r_e)$, for each galaxy;  and in Fig.~\ref{fig:ELL_MASS_COADDED}
the coadded mass distributions for the high-$L$ and the low-$L$ subsamples
that can be very well reproduced (solid lines) by a two-component mass model
which includes a luminous Hernquist spheroid and a DM halo given by
eq.(\ref{eq:HALO_VEL}). The resulting fit is excellent (see
Fig.\ref{fig:ELL_MASS_COADDED}) when $\beta$, the luminous mass fraction
inside $R_{opt} \simeq 2\, r_e$, scales with luminosity as in disk systems,
and the halo core radius $a$ (expressed in units of $R_{opt}$) scales as
\begin{equation} 
a ~=~  0.8 ~\biggl( {L \over L_*} \biggr)^{0.15}\,.
\end{equation}
The central density of the DM halo, $\rho(0) \simeq 1.5 \times 10^{-3}
(1-\beta)\, M(R_{opt})/ R_{opt}^3\, (1+a^2)/a^2$, is $3-5$ times larger
than the corresponding densities in spirals, and it scales with luminosity
in a similar way to the case of spirals. These results confirm those of
the pioneering work of Bertola et al. (1993). 

Finally, in Fig.~\ref{fig:ELL_ML}, we show the derived $(M/L)_\star$ as a
function of color (right) and of luminosity (left).  The stellar
mass-to-light ratios are consistent with the predictions of population
synthesis models (e.g., Tinsley 1981) and are compatible with {\it (a)}
the $M/L \propto L^{0.2}$ relation deduced from the tilt of the
Fundamental Plane (e.g. Diorgovski \& Santiago 1993), and {\it (b)} with
van der Marel's (1991) result $(M/L)_\star \propto L^{0.35 \pm 0.05}$ (in
the $R$-band), the only work to-date in which a dynamical derivation of
stellar $M/L$ ratios has been obtained for a large sample of ellipticals.

\section{Dwarf Irregulars}

\begin{figure}
\par
\centerline{\vbox{
\psfig{figure=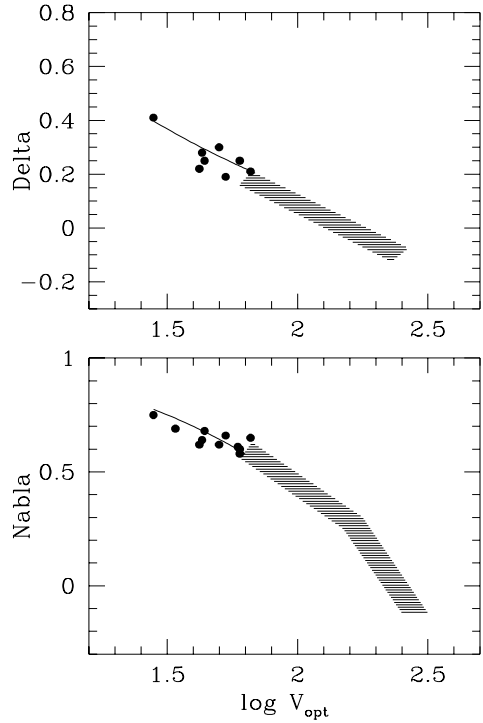,height=9cm,width=4cm} }}
\par
\caption {Inner and outer RC gradients for a sample of dIrr's from the 
literature. The shaded areas represent the loci populated by spirals.
The solid lines are the prediction from the `spiral' mass model. }
\label{fig:DIRR_NABLA_DELTA}
\end{figure}

The stellar component of these disk systems is distributed according to an
exponential thin disk as in spirals (Carignan \& Freeman 1988). At high
luminosities, dIrr's barely join the low-$L$ tail of spirals; at low
luminosities, they reach down to $\sim 10^{-3} L_*$.  dIrr's have very
extended HI disks: high-quality RCs can be then measured out to $2 \,
R_{opt}$ (e.g. C\^ot\'e et al. 1997; Swaters 1997).  The dark matter
presence is apparent when we plot (see Fig.~\ref {fig:DIRR_NABLA_DELTA})
the inner and outer RC gradients, $\nabla$ and $\delta$, as functions of
velocity. Immediately, we realize that the DM fraction is overwhelming:
$\nabla >> -0.3$, gently continuing, at smaller $V_{opt}$, the trend of
low-luminosity spirals.  The extent and the quality of these RCs permit
reliable determinations of the halo parameters by working out the mass
model that best reproduces the RC shapes, described by the quantities 
$\nabla$ and $\delta$. 
   
\begin{figure}
\par
\centerline{\vbox{
\psfig{figure=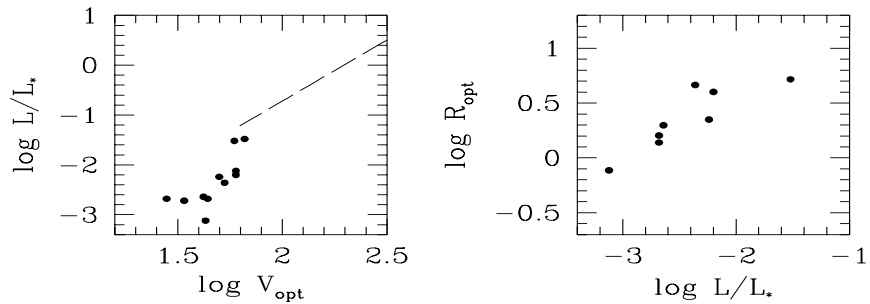,height=3.5cm,width=5cm} }}
\par
\caption {{\it Left:} the TF relation for dIrr's; the dashed line indicates 
the spirals TF (from PSS96). 
{\it Right:} the luminosity--radius relation for dIrr's.} 
\label{fig:DIRR_TF_LUM_RAD}
\end{figure}

\begin{figure}
\par
\centerline{\vbox{
\psfig{figure=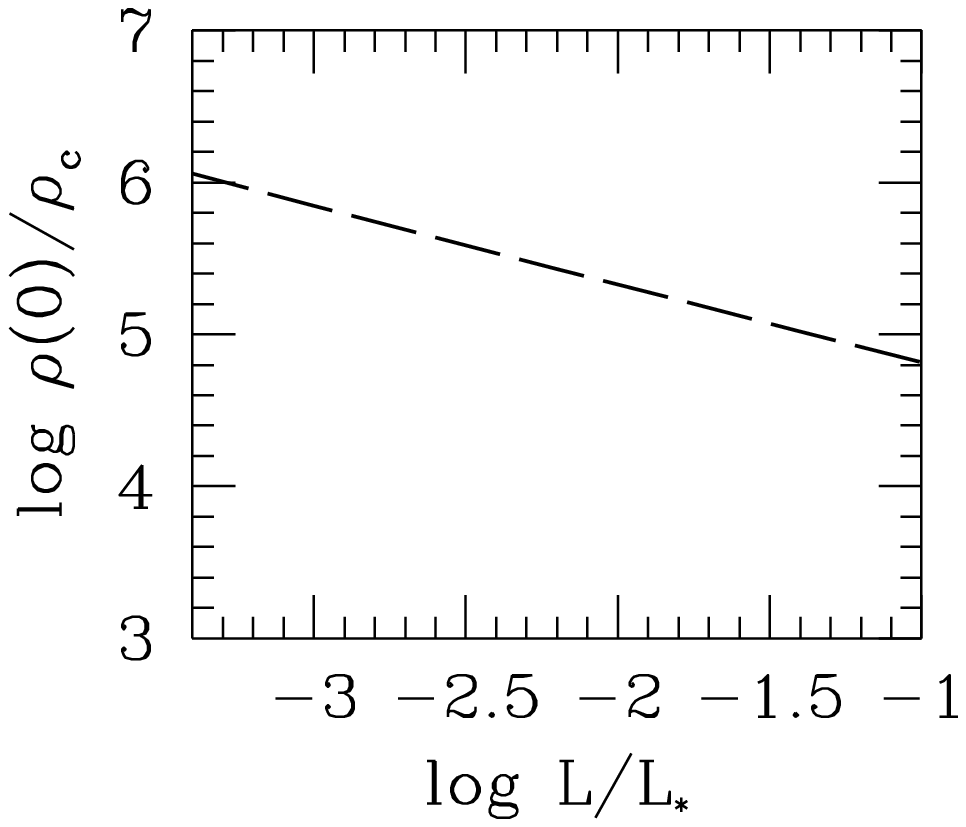,height=5.5cm,width=5.5cm} }}
\par
\caption { dIrr halo central density as a function of 
luminosity.} 
\label{fig:DIRR_DENS_LUM}
\end{figure}

Such a mass model includes   
a disk, and a dark halo with `spiral' mass profile [see 
eq.(\ref{eq:HALO_VEL})].
We recall that $x \equiv
R/R_{opt}$, $a$ is the halo core radius in units of $R_{opt}$, and $\beta$ is
the visible mass fraction, also inclusive of the gas content,
at $x=1$.  The outer slope $\delta \equiv V(2)/V(1)-1$ is related to the
dark and visible matter structural parameters by (see PSS96): 
\begin{equation}
\delta ~=~ \biggl[(1+\delta_\star)^{2} \beta ~+~ (1-\beta) {4 \, (1+a^2) 
\over 4+a^2} ~-~ 1 \biggr]^{1/2}\,,
\label{eq:EXT_DIRR}
\end{equation}
while the inner slope $\nabla$ is related by
\begin{equation}
\nabla ~=~ \beta \, \nabla_\star ~ +~ (1-\beta) \, \nabla_h\,,
\end{equation}
with $\nabla_\star \sim -0.2$ including also the HI content. In detail, we
aim to reproduce both the $\nabla$--log$V_{opt}$ and
$\delta$--log$V_{opt}$ relationships, by means of the above-described mass
model [from the definition:  $\nabla_h= {a^2\over (1+a^2)} (0.86+0.5/a)$].
The contribution of the baryonic disk (stars + gas) to the circular
velocity is roughly constant with radius (e.g., Carignan \& Freeman 1988):
the decrease of the stellar contribution outside $R_{opt} ~=~ 5 \times (
{L \over 0.04 \, L_*})^{0.45}$ is counterbalanced by the increase of
the gas contribution, and therefore $\delta_\star \simeq 0$.  An excellent
agreement between the model and observations is reached when: 
\begin{equation}
a ~=~ 0.93 \times \biggl( { V_{opt} \over 63~ {\rm km~s}^{-1}} \biggr)^{-0.5 
}\,, \label{eq:DIRR_CORE_RADIUS}
\end{equation}
and 
\begin{equation}
\beta ~=~ 0.08 \times \biggl( { V_{opt} \over 63~ {\rm km~s}^{-1} } 
\biggr)^{1.2}\,. 
\label{eq:DIRR_BETA}
\end{equation}

A luminosity--velocity relation for dIrr's, shown in
Fig.~\ref{fig:DIRR_TF_LUM_RAD}, continues down to M$_B=-14$ the TF
relationship for spirals, allowing us to write the halo structural
parameters as a function of galaxy luminosity: 
\begin{equation}
V_{opt} ~=~ 63 ~ \biggl({L\over 0.04 \, L_*}\biggr)^{0.16} ~~ {\rm km~~s}^{-1}
\label{eq:DIRR_TF}
\end{equation}
 
The central DM density computed by means of eqs.(\ref{eq:DIRR_BETA}),
(\ref{eq:HALO_VEL}), (\ref{eq:DIRR_CORE_RADIUS}), is plotted in
Fig.~\ref{fig:DIRR_DENS_LUM}. We realize that dwarf galaxies have the
densest halos, continuing the inverse trend with luminosity of spirals. Finally,
these objects are the darkest in the Universe: $\beta$ continues to
decrease at lower luminosities down to $\sim 10^{-2}$. Notice that dIrr
halos may have ``larger'' core radii, in units of $R_{opt}$, inverting the
trend with luminosity detected in larger objects. 

These results are in good agreement with the best-fit mass models of
individual RCs (Puche \& Carignan 1991; Broeils 1992; C\^ot\'e et al. 
1997; Swaters 1997). They in fact show that DM halos, with large core radii
$> R_D$, completely dominate the mass distribution of these objects. In
greater detail, C\^ot\'e (1995) and C\^ot\'e et al. (1997) compiling previous work
find that, at $x \simeq 2$, $M_{dark}/M_{bar}= 1 - (M_B+20)$ and that
central densities increase by a factor $\sim 10$ from M$_B =-20$ and M$_B=
-14$. Both results are in good agreement with the present work. 

Let us point out that dIrr's, although having a negligible amount of
light, do have quite a large mass: $\sim 8 \times 10^{10} (L/L_{\rm
max})^{1/3}$ with $L_{\rm max}=0.04\, L_*$ being the maximum luminosity 
observed in this family.

\section{Dwarf Spheroidals}

Dwarf ellipticals/spheroidals (dSph's) are the faintest observed galaxies
in the Universe and the least luminous stellar systems. Yet they
represent the most common type of galaxy in the nearby universe (e.g.,
Ferguson \& Binggeli 1994). 

Given the low surface brightnesses involved, the main kinematical quantity
tracing the gravitational potential, i.e. the velocity dispersion, can be
determined by measuring redshifts of individual stars. Ever since early
measurements of dSph velocity dispersions, high $M/L$ ratios were derived
implying large amounts of DM in these objects (Faber \& Lin 1983).
Kormendy (1988) and Pryor (1992) have shown that dSph's are DM dominated
at all radii:  core fitting methods (Richstone \& Tremaine 1986) yielded
central DM densities of $0.1 M_\odot {\rm pc}^{-3}$ (i.e., overdensities
of $10^7$), a factor 10--100 larger than the stellar ones. 

\begin{figure}
\par
\centerline{\psfig{figure=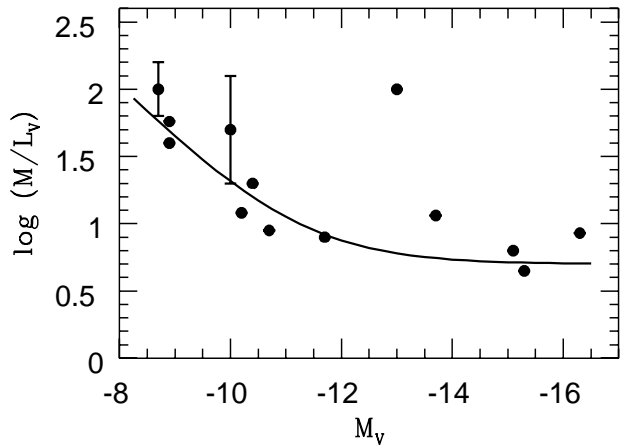,height=6.2cm,width=7cm}}
%\centerline{\psfig{figure=dSph.ps,height=7cm,width=7.7cm}}
\par
\caption { The $M/L$ ratios of dIrr's: the observational data (from Mateo 
1997: circles) vs. our prediction from the `spiral' mass model (line) } 
\label{fig:DSPH}
\end{figure}

However, only recent kinematic studies (Armandroff et al. 1995; Hargreaves
et al. 1994, 1996;  Ibata et al. 1997; Mateo 1994; Queloz et al. 1995;
Vogt et al. 1995) have gathered a suitable number ($\magcir 10$) of
galaxies with central velocity dispersion derived by repeat measurements
of motions of $>> 10$ stars. Mateo (1997), analysing this observational
data, has shown that the central $M/L$ increases with decreasing galaxy
luminosity (see Fig.\ref{fig:DSPH}), implying, even in the innermost
regions, the presence of a dark component whose importance increases with
decreasing luminosity.  We fit this $M/L$--$L$ relationship by modelling
the mass of these galaxies with {\it (i)} a luminous mass component with
$M_\star=5\ L_V$ (as Mateo 1997) and {\it (ii)} a dark halo with profile
given by eq.(\ref{eq:HALO_VEL}). Then, also these objects have a `spiral'
dark-to-luminous mass ratio.  For $r<<r_e$ the mass profile of a Hernquist
spheroid is very similar to that of a `sphericized' Freeman disk, for
which we can write $L(x)=L_V [1-(1+3.2x)e^{-3.2\,x}]$. We obtain the
model's central $M/L$ ratios from $\lim_{x \rightarrow 0}
[M_h(x)+M_\star(x)]/L(x)$ (see Fig.\ref{fig:DSPH}). The agreement between
the model and observations is striking, although {\it (a)} we have
extrapolated luminosities and radial distances by a factor of $\sim 100$
and {\it (b)} the model has no free parameter. As a consequence, we find
$M_h \propto L_V^{1/4}$, which is indistinguishable from Mateo's best fit,
$M_h \sim 2 \times 10^{7}M_\odot$ independent of luminosity, and in
qualitative agreement with pioneering studies of the dSph Fundamental
Plane (Ferguson \& Binggeli 1994). 

The present data, referring to central regions, do not allow one to derive
the run of $a$ with luminosity and then evaluate the central halo density.
Taking a tentative value for $a \sim R_c$, the King radius, yields central
overdensities $10^{6}-10^{7}$ which make these galaxies among the
densest of the Universe, as is obvious from the fact that these tiny
galaxies withstand, at a distance of few tens of kpc, the gravitational
tides of large galaxies like M31 and our own (Gallagher et al. 1997).

\section{The structure of dark halos}  

In the previous section we have shown that the structure of dark halos
around galaxies has a universal character: a specific functional form,
with two mass dependent-parameters, describes the density distribution of
dark halos at any radius, in all galaxies of every Hubble type. Here, we
discuss the basic properties of such a DM distribution and the relevant
aspects of the interplay between dark and luminous matter.  These can be
divided into {\it (a)} universal properties, that is, not depending on
Hubble type, and {\it (b)} morphological properties which do depend on
Hubble type. 

\subsection{Universal properties}

From the kinematics of a very large number of disk and spheroidal systems,
there unquestionably emerges a one-to-one relation between a luminous galaxy
and a massive dark halo.  This is the rule: a hypothetical galaxy found
with no dark halo should be considered as a peculiar exception. Galaxies
lie within large self-gravitating dark halos: more specifically, a
disk/spheroid of size $R_{opt}$ and mass $M_\star$, should be actually seen as
embedded in a dark halo of radius $\sim R_{200} \sim 15 \,R_{opt}$ and
mass $M_{200} =3 \times 10^{12} ({M_\star \over 2\times 10^{11}
M_\odot})^{0.4}M_\odot$. 
\medskip

\noindent
$\rhd$ The luminous matter, not surprisingly considering the dissipational
collapse it has experienced, is more concentrated, by a factor $\sim
10-15$, than the dark matter. This explains why the former nearly always
dominates the inner regions of galaxies, where the luminous-to-dark matter
density ratio is always $>> 10^{4} \Omega_b \beta >> 1$.  From the
galaxy center out to $ R_{opt}$, the fraction of DM goes from $0\%$ up to
$30\% - 70\%$. Thus, all across the region where the baryonic matter
resides, the dark and luminous components are well mixed, except in
very-low-luminosity galaxies, dominated by the dark matter at all radii.
Thus, two common (and competing) ideas, according to which either {\it i)}
dark matter is the main component inside $R_{opt}$ or {\it ii)} dark
matter is important only where the stellar distribution ends, are both
ruled out by observational evidence. In the same way, the very concept of
mass-to-light ratio retains its physical meaning only if the radius at
which a value is derived is specified. 
\medskip

\noindent
$\rhd$ DM halos have core radii, i.e. regions of width $\sim R_{opt}$,
where the DM density remains approximately constant. Let us stress that
the core is apparent in every galaxy of every Hubble type, not only in
dIrr's (see Moore 1994). Actually, this region is larger in larger
galaxies, both in physical and in normalized units. The existence of core
radii makes the central density of a DM halo a well-defined, physically
meaningful quantity, and it implies that DM halos, even though arising from
scale-free perturbations, do actually develop a scale, related with the
half-light scale. The well proven existence of core radii, furthermore,
rules out all halo models with a prominent central cusp or with a hollow
core.  
\medskip
 
\begin{figure}
\par
\centerline{\psfig{figure=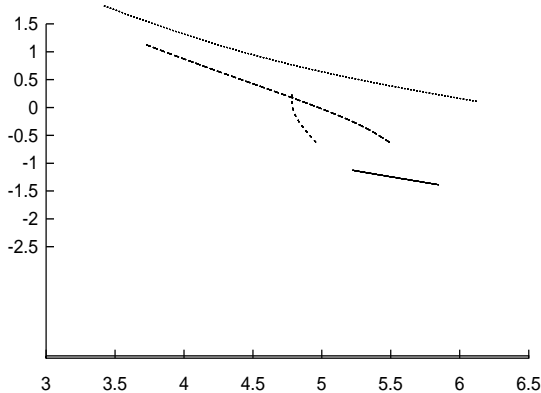,height=6.5cm,width=5cm,angle=-90}}
\par
\caption { Luminous-to-dark mass ratio vs. halo central density diagram. 
Always, higher DM fractions correspond to higher densities. 
(Dotted curve: E's; long-dashed curve: spirals; short-dashed curve: LSBs; 
solid curve: dIrr's.)} 
\label{fig:DENS_ML} 
\end{figure}

\noindent
$\rhd$ The DM central densities range through about 3 orders of magnitude,
inversely correlated with galaxy luminosity (see Fig.\ref{fig:DENS_ML}),
consistently with hierarchical scenarios of galaxy formation in which
smaller objects form first. 
\medskip

\noindent
$\rhd$ Proto-galaxies are likely to start their collapse with the
same baryon fraction $\Omega_b$. However, in present-day galaxies this
quantity strongly depends on the galaxy luminosity (halo mass), ranging
from $\sim \Omega_b$ at high masses $\magcir 10^{11} M_\odot$, to $10^{-4}
\Omega_{b}$ at the lowest mass $\sim 10^{8} M_\odot$. The efficiency of
retaining the primordial gas and   transforming  it in stars is then a
strong function of the depth of the (halo) potential well. 
\medskip

\begin{figure}
\par
\centerline{\psfig{figure=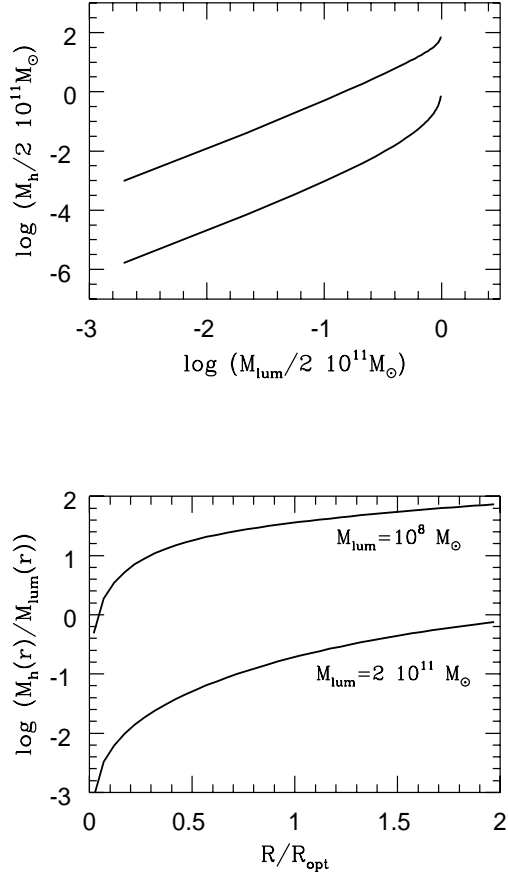,height=12cm,width=6cm}}
\par
\caption { The dark-to-luminous mass ratio for high and low values of the
stellar mass ($M_\star=2 \times 10^{11}M_\odot$ and $M_\star = 10^8
M_\odot$, respectively) as a function of normalized radius {\it (bottom)};
within a specified radius, DM mass vs. luminous mass {\it (top)}.}
\label{fig:GAMMA} 
\end{figure}

\noindent
$\rhd$ The range in luminosity among galaxies, $>3$ orders of magnitude,
is much wider than that of halo masses, which spans through $<2$ orders of
magnitude. This implies that the global mass-to-light ratios of galaxies
decrease with increasing halo mass as ${M_{200}\over L } \propto
M_{200}^{-1}$. According to the above, considering all galaxies as
having the same total mass, is not as bad as assuming that their masses
and luminosities are directly proportional.  In this light, the persistent
habit in many cosmological studies of assuming $M_{\rm halo}/L=const$ should
be avoided, if possible.  
\medskip

\noindent
$\rhd$ In every galaxy and at any radius, the distributions of dark and
luminous matter are coupled: the luminous matter knows where the dark
matter is distributed and viceversa. The coupling is universal, in that it
is essentially independent of the Hubble type of the galaxy. At a (normalized)
radius $x=R/R_{opt}$ the mass ratio takes the form: 
$$
{M_h(x) \over M_\star (x)} ~=~ 0.16 ~\biggl({M_\star \over 2\times 
10^{11} 
M_\odot}\biggr)^{3/4} ~ \biggl[1+3.4 \,\biggl({M_\star \over 2\times 10^{11} 
M_\odot}\biggr)^{1/3}\biggr] ~~~~\times ~~~~~~~~~~~~~~~~~~~~~~~
$$
\begin{equation} 
~~~~~~~~~~~~~~~~~~ { ~x^2 \over 
\biggl[ x^2 ~+~ 3.4\, ({M_\star\over 2 \times
 10^{11} M_\odot})^{1/3} \biggr] \biggl[1-(1+3.2\,x)\, e^{-3.2\,x}
\biggr]}\,.
\end{equation}

In Fig.(\ref{fig:GAMMA}) we show the radial dependence of the
dark-to-luminous mass ratio for the highest and lowest stellar masses, $2
\times 10^{11} M_\odot$ and $10^8 M_\odot$, and the dark matter inside a
fixed radius $R$ as a function of the luminous matter inside that radius. 
The dark-luminous coupling can be quantified by noting that, where the
luminous mass is located ($R \mincir 2\,R_{opt}$), over five orders of
magnitude in mass and independently of the total stellar (or halo) mass,
$M_\star\propto M_h^{2/3}$; this relationship breaks down where the
stellar distribution converges. This interplay is the imprint of the late
stages of the process of galaxy formation and rules out the concept of
``cosmic cospiracy''.  On the theoretical side, we emphasize that it is
difficult to envisage how such a structural feature may arise in scenarios
radically different from a (CDM-like) bottom-up. 

\subsection {Morphological dependence}

\begin{figure}
\par
\centerline{\psfig{figure=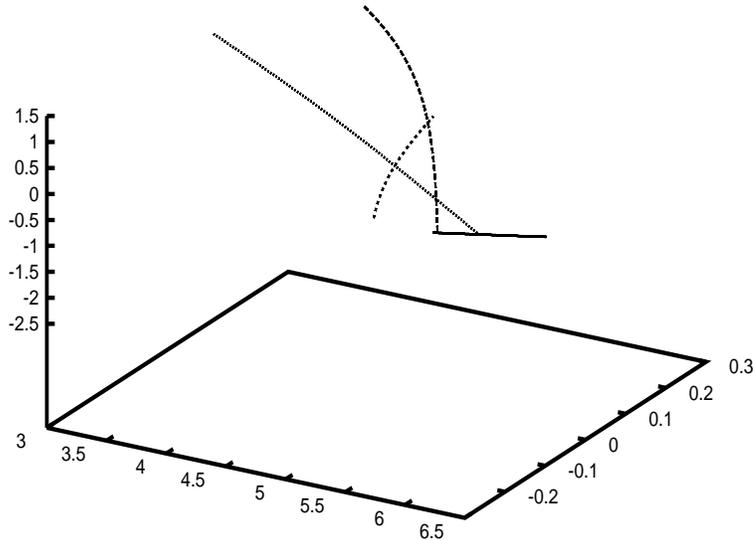,height=7cm,width=5cm,angle=-90}}
\par
\caption { The loci populated by the families of galaxies in the
luminous-to-dark mass ratio at $R_{opt}$, halo central density, and (halo
core)-to-optical radius ratio space ($z/x/y$).  (From left to right, we 
encounter ellipticals, spirals and LSBs, and dIrr's.)}
\label{fig:GRANROSARIO}
\end{figure}

While the universal behaviour of the mass distribution of dark halos  relates to
the cosmological properties of DM, any Hubble type dependence of the
distribution of the luminous and dark matter characterizes the late
processes of galaxy formation, such as  transfer of angular momentum, the
efficiency of the star formation and its feed-back on galaxy structure. 
Remarkably, a small number of  structural quantities allow 
one to describe the gross features
of the  morphological/luminosity dependence of the DM distribution and of the
interplay with the luminous matter. These are: the central DM density
$\rho_0$, the core radius $a$ (in units of $R_{opt}$) and the
luminous-to-dark mass ratio evaluated at the optical radius. For 
each Hubble Type, the above structural paramenters are all functions of
luminosity and then correlate among themselves.  In
Fig.(\ref{fig:GRANROSARIO}) we show, for each Hubble type, the curves
generated in the space $\rho_0,\, a,\, {M_h \over M_\star}|_{x=1}$ by the 
variation of luminosity. We note that:  
\medskip

\noindent 
$\circ$ the DM structural parameters and the connection with the luminous
matter show a strong continuity when passing from one Hubble Type to
another. This happens despite the fact that both the distribution and the
global properties of the luminous matter show strong morphological
discontinuities; 
\medskip

\noindent 
$\circ$ dwarf galaxies, the densest galaxies in the Universe, are also
completely dark-matter dominated: however, their low baryon content just
smoothly continues downwards the dependence with galaxy mass followed by larger
galaxies. As this continuity extends to all other structural properties,
the tendency to theoretically investigate these objects separately
from ``normal'' galaxies could be misleading; 
\medskip

\noindent 
$\circ$ spirals show the largest range in dark-to-luminous mass ratios and
central densities, indicating that the occurrence of this morphological
type is independent of the structure/evolution parameters considered here.
This is probably because the main factor responsible for the formation
of disk systems, i.e.  the content of angular moment, is likely to be
independent of halo mass; 
\medskip
   
\noindent 
$\circ$ LSB galaxies are significantly less dense than normal spirals.  As
this is the case for both the dark and luminous components, the
fractional amount of dark matter is not affected. This ratio, as well as
the size of the halo core radius, depends on galaxy luminosity as in HSB
spirals. LSBs, instead, are clearly distinguished by having both lower
$\rho_0$ and lower stellar mass-to-light ratios. This suggests that the
differentiation between HSB and LSB galaxies is due to different initial
conditions (e.g., content of angular momentum, epoch of formation) rather
than being developed during the late stages of formation;  
\medskip

\noindent $\circ$ ellipticals, considered as luminous spheroids, are well
characterized objects, evidently very different from disk systems.
However, in the structural parameter space, E and S galaxies are
contiguous, the main difference being that the former are more
concentrated in both the dark and luminous components. Combined with the
evidence that the dark halos of ellipticals have smaller core radii (both in
normalized and in physical units), this morphological property may be due
to a deeper baryonic infall in the halo potential well.  
\bigskip

\noindent 

{\bf Acknowledgements.} We thank Erwin de Blok for sending his LSB
rotation curve data. We also thank Lynn Matthews for communicating data
prior to publication. We acknowledge John Miller for carefully reading the
manuscript. Finally, special thanks are due to all the participants to the
Sesto DM1996 conference, without whom this paper would not have been
possible.

\newpage

\section{Appendix}

In this paper we have used data coming from many different sources. In 
detail:
\medskip

\noindent
$\Box$ {\sl Spiral Galaxies.} See references quoted in PSS96.
\bigskip

\noindent
$\Box$ {\sl LSB Galaxies.} We have used all the objects in the de Blok et
al.  (1996) sample, except for: F567-2, F577-V1, F579-V1 (asymmetric
velocity arms), F571-V2 (optical scalelength missing), and F564-V3
(rotation curve missing). 
\bigskip

\noindent
$\Box$ {\sl Elliptical Galaxies.} The sources are:
\medskip

\noindent
N 720: Buote \& Canizares 1994, ApJ, 427, 86;

\noindent
N1052, N2974, N4278, N5077, N7097, I2006: 
   
\noindent
\hskip 1.3cm Bertola et al. 1993, ApJ, 416, L45 (and refs. therein);

\noindent
N 1453: Pizzella 1997, private communication;

\noindent
N 1399: Ikebe et al. 1996, Nature, 379, 427;

\noindent
N 4697: Dejonghe et al. 1996, A\&A, 306, 363;

\noindent
N 5128: Hui et al. 1995, ApJ, 449, 592;

\noindent
N 6703: Saglia et al. 1997, this volume. 
\bigskip

\noindent
$\Box$ {\sl Dwarf Irregular Galaxies.} The sources are:
\medskip

\noindent
DDO 154:  Carignan \& Freeman 1988, ApJ, 332, L33;

\noindent
UGC 442, E381-G20, DDO 161, E444-G84: C\^ot\'e 1997, this volume;

\noindent
DDO 170: Lake et al.1990, AJ, 99, 547;

\noindent
I3522, U7906: Skillman et al. 1987, A\&A, 185, 61;

\noindent
DDO 175: Skillman et al. 1988, A\&A, 198, 33;

\noindent
UGC 12732, DDO 9: Swaters 1997, this volume.
\bigskip

\noindent
$\Box$ {\sl Dwarf Spheroidal Galaxies.} The sources are Mateo (1997) and 
references therein. 
\medskip

\end{document}